\documentclass[
    ,final            % use final for the camera ready runs
  ]
  {aipproc}

\layoutstyle{8x11double}

\begin{document}

%%\title{An atomically precise Si(331)-(12$\times$1) }
\title{Atomically precise Si(331)-(12$\times$1) surfaces}

\classification{68.35.bg } \keywords {Silicon surface
reconstructions}

\author{Corsin Battaglia}{
  address={Institut de Physique, Universit\'e de Neuch\^atel,
2000 Neuch\^atel, Switzerland} }

\author{Claude Monney}{
  address={Institut de Physique, Universit\'e de Neuch\^atel,
2000 Neuch\^atel, Switzerland} }

\author{Cl\'ement Didiot}{
  address={Institut de Physique, Universit\'e de Neuch\^atel,
2000 Neuch\^atel, Switzerland} }

\author{Eike Fabian Schwier}{
  address={Institut de Physique, Universit\'e de Neuch\^atel,
2000 Neuch\^atel, Switzerland} }

\author{Nicolas Mariotti}{
  address={Institut de Physique, Universit\'e de Neuch\^atel,
2000 Neuch\^atel, Switzerland} }

\author{Michael Gunnar Garnier}{
  address={Institut de Physique, Universit\'e de Neuch\^atel,
2000 Neuch\^atel, Switzerland} }

\author{Philipp Aebi}{
  address={Institut de Physique, Universit\'e de Neuch\^atel,
2000 Neuch\^atel, Switzerland} }

\begin{abstract}
Si(331)-(12x1) is the only confirmed planar silicon surface with a
stable reconstruction located between (111) and (110). We have
optimized the annealing sequence and are able to obtain almost
defect free, atomically precise surface areas approaching micrometer
dimensions. The unprecedented perfection of the surface combined
with its pronounced structural anisotropy makes it a promising
candidate to serve as template for the growth of self-assembled
one-dimensional nanostructure arrays.
\end{abstract}

\maketitle

%%%%%%%%%%%%%%%%%%%%%%%%%%%%%%%%%%%%%%%%%%%%
%% MAINMATTER
%%%%%%%%%%%%%%%%%%%%%%%%%%%%%%%%%%%%%%%%%%%%

%\section{Introduction}

\begin{figure}\centering
 \resizebox{1.\columnwidth}{!}
{
\includegraphics{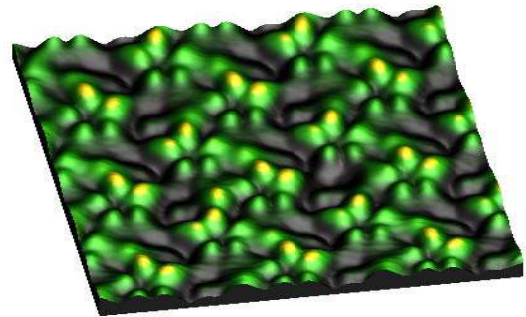}}%
\caption{\label{fig:STM331}(Color online) 3D rendered
high-resolution STM topography of the Si(331)-(12$\times$1)
reconstruction. Bias voltage 2.0 V, set-point current 0.06 nA, size
7 nm $\times$ 5 nm, temperature 77 K. }
\end{figure}

The fabrication of atomically precise nanostructure arrays on
semiconductor surfaces remains a challenging task. A narrow size
distribution and controlled positioning are essential in view of
applications, since fluctuations of only a few atoms may
substantially alter the electronic and optoelectronic properties.
Spontaneous self-assembly represents a promising bottom-up approach
allowing production of nanostructures in a massively parallel
fashion. However, the uniformity often suffers from relatively large
variations and the
spatial arrangements are hard to control.\\
A prototypical example for atomically precise self-assembled
nanostructures are semiconductor surface reconstructions, which form
not only on clean surfaces \cite{Battaglia08b}, but may also be
stabilized by foreign adsorbate species
\cite{Battaglia07,Battaglia08c}. They are ideal templates for the
growth of more complex atomically precise structures such as magic
clusters exhibiting enhanced stability at certain magic sizes.
Atomically precise periodic surfaces, such as the
Si(111)-(7$\times$7) reconstruction, have been used successfully to
grow
perfectly ordered arrays of such identical-size clusters \cite{Li02}. \\
Here we focus on the Si(331)-(12$\times$1) reconstruction, whose
surface normal is located 22$^o$ away from the (111) direction
towards (110). Based on scanning tunneling microscopy (STM) images
of unprecedented resolution, we were recently able to resolve for
the first time double rows of silicon pentamers on this surface (see
Fig. \ref{fig:STM331}) \cite{Battaglia08}. Combining our STM images
with low energy electron diffraction data and first-principles total
energy
calculations allowed us to propose a new structural model. \\
Surface preparation is crucial if one wants to study structural and
electronic properties at the atomic scale. Here we describe in more
detail our optimized sample preparation process. Using an automated
system, Si(331)-(12$\times$1) surfaces of very high quality are
prepared routinely.\\
\begin{figure}[t]
\centering
\includegraphics[draft=false]{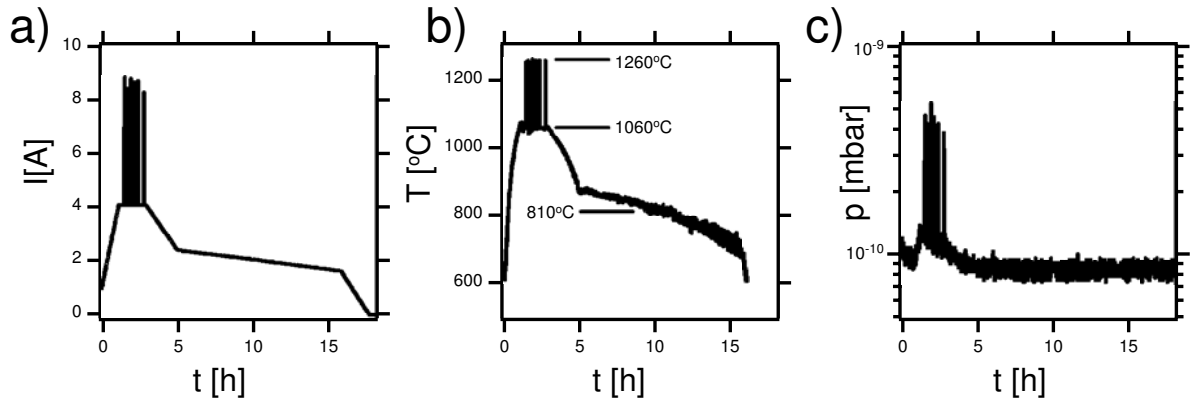}%
\caption{\label{fig:Chapter1_Preparation} Typical preparation cycle
for a Si(331) surface. a) current vs time, b) temperature vs time,
c) pressure vs time.}
\end{figure}
Boron doped Si(331) samples from Crystec with a resistivity of
0.1-30 $\Omega$ cm were used in this work. The reconstruction of the
Si(331) surface is highly sensitive to small amounts of Ni
contamination changing the periodicity from (12$\times$1) to
(13$\times$1). To avoid Ni contamination, any contact of silicon
with stainless steel must be avoided. Silicon samples and the Mo
sample holders are exclusively handled with teflon tweezers and Mo
tools. All samples are first cleaned for 5 minutes in an acetone
ultrasound bath followed by an ethanol ultrasound bath for another 5
minutes. Coming out of the ethanol bath, the samples are dry blown
with clean
nitrogen gas, mounted on the Mo holder and transferred into vacuum.\\
During the first preparation step in vacuum, the sample is degassed
by passing a direct current through the sample. In our setup a
pressure signal from the gauge controller is sent into a feedback
loop which controls the sample current while keeping the pressure
below $5\times 10^{-10}$ mbar. Once the sample temperature attains
1060$^o$C, the sample is flashed repeatedly to 1260$^o$C. The ramp
from 1060$^o$C to 1260$^o$C is performed within 2 seconds, the
cool-down back to 1060$^o$C is completed within 10 seconds during
the first flashes in order to keep the pressure during the entire
flash below $2\times 10^{-9}$ mbar. The cool-down period during the
final flashes is 30 s. At 1060$^o$C single steps on Si(111) have
been reported to be stable \cite{Viernow98} and we find it to be an
ideal base temperature to perform flashing to higher temperature,
since it is above the transition temperature for all known silicon
surface reconstructions. The stress on the wafer, induced by the
fixation to the sample holder, should be minimized to
allow the sample to expand freely upon heating.\\
The cool-down sequence which follows is crucial for the quality and
long range order of the surface reconstruction. Ideally one would
like to keep the sample for a certain period at the transition
temperature, where the surface reconstruction forms, to allow long
range order to be established. However, transition temperatures
reported in the literature are usually accurate to within only
20$^o$C to 30$^o$C. Additional uncertainties arise due to the
difficulty to obtain an absolute calibration for our own pyrometer
which works at a single wave length. Uncertainties due to value of
the selected emissivity should also not be forgotten. All our
measurements on silicon
surfaces were carried out using an emissivity $\epsilon$ of $0.65$.\\
Our cool-down strategy is simple, but yet highly effective. Instead
of annealing the sample at a fixed temperature, we start the
annealing sequence at about 50$^o$C above the transition
temperature, which was reported to be at 810$^o$C \cite{Wei91}, and
slowly cool the sample down to a temperature of about 50$^o$C below
the transition temperature typically within 12 h. The slow cool-down
rate guarantees that the sample spends enough time close to the
transition temperature to optimize its long range order. Sample
preparation is completed by a cool-down of 2 h down to room
temperature. With this method we are able to obtain almost defect
free, atomically precise Si(331)-(12$\times$1) surfaces over areas
approaching micrometer dimensions.\\
Fig. \ref{fig:Chapter1_Preparation} shows the preparation cycle for
the Si(331) surface. For process control purposes it is useful to
record not only the current a) and the temperature b) of the sample
as a function of time but also the pressure c) during the entire
preparation cycle. The three automated steps of the preparation
cycle, degassing, flashing and cool-down including annealing, can be
clearly distinguished in all three traces. Note that the
Si(331)-(12$\times$1) reconstruction forms at a temperature of
810$^o$C, which lies
approximately in the middle of the annealing temperature range.\\
In summary we have described the preparation process for obtaining
high quality Si(331)-(12$\times$1) surfaces for which we recently
proposed a new structural model.

%\begin{theacknowledgments}
This work was supported by the Fonds National Suisse pour la
Recherche Scientifique through Division II and the Swiss National
Center of Competence in Research MaNEP.
%\end{theacknowledgments}
\enlargethispage{1cm}

\bibliographystyle{aipproc}   % if natbib is available

\bibliography{Si331_ICPSII}

\end{document}